\documentclass[%
 reprint,
superscriptaddress,
 amsmath,amssymb,
 aps,
]{revtex4-2}
\usepackage{amsmath}
\usepackage{graphicx}
\usepackage{dcolumn}
\usepackage{bm}

\begin{document}

\preprint{APS/123-QED}

\title{ High Finesse Cavity with Anapole-Assisted Resonant Subwavelength Particle Mirror}

\author{Zheng Xi}\email{z.xi@tudelft.nl}
\affiliation{Department of Optics and Optical Engineering, University of Science and Technology of China}
\affiliation{Department of Imaging Physics, Optica Research Group, Delft University of Technology}
\author{H.P. Urbach}%
\affiliation{Department of Imaging Physics, Optica Research Group, Delft University of Technology}%

\date{\today}

\begin{abstract}Strong light interaction with a subwavelength object has been a long pursuing goal with difficulties mainly arising from the diffraction limit.  We propose a high finesse cavity with one mirror made of a subwavelength resonant particle as a platform to enhance this interaction.  High quality eigenmode solutions  are  obtained  for  such  a  highly  non-paraxial  cavity  with  a  very  high  field  concentration at the particle. The eigenmode solutions interact with the small particle in a more general way than by the electric dipole approximation.  With the help of the anapole excitation in the dipole term, the particle is designed to scatter like a pure magnetic quadrupole, and in this way, it has a near-unity reflectivity when used as mirror for the strongly focused field of the eigenmode.  Light-matter interactions at the subwavelength scale can be greatly enhanced due to the small size of the particle and the high finesse of the cavity, which can be potentially interesting for applications in nano optics, quantum optomechanics, nonlinear optics, and subwavelength metrology beyond the electric dipole approximation.
\end{abstract}
\maketitle


Cavities are important tools to enhance light-matter interactions. In particular, a high finesse cavity with a subwavelength sized mirror can be of great interest because of the very high field concentration at the small mirror and the greatly enhanced interaction strength due to the high finesse. This can, for example, lead to enhanced optomechanical interaction\cite{aspelmeyer2014cavity} since the effect of radiation pressure can be maximized with the smallest mirror and the highest cavity finesse, which has important applications ranging from high precision metrology\cite{abbott2016observation}, optical cooling\cite{delic2020cooling,kleckner2006sub}, massive quantum super-positions\cite{marshall2003towards}, nonlinear interactions\cite{fabre1994quantum} and even optomechanics at the single-photon level\cite{nunnenkamp2011single}.

While there is a tremendous amount of effort in pursuing the design and realization of such a cavity\cite{bettles2016enhanced,hetet2011single,heugel2010analogy,rui2020subradiant,sondermann2015photon,zumofen2008perfect}, there are still two substantial obstacles. First, in conventional cavities, stable eigenmodes are paraxial fields  with slowly varying lateral field amplitude.  
This paraxial theory is no longer valid for a cavity with a subwavelegnth mirror. It is generally thought that such a cavity always has large diffraction losses. As a consequence, current mirrors are often made much larger than the wavelength to stay within the paraxial regime as illustrated in Fig. 1(a). The second obstacle is that small scatterers are in general not good reflectors. Indeed, if the scattering can be approximated by that of a point dipole or by Rayleigh scattering, the scattered power scales as ${r^6}/{\lambda ^4}$ with $r$ being the averaged size of the particle and $\lambda$ the wavelength. For $r\ll\lambda$, the interaction is weak which makes it very hard to realize a high finesse cavity.

In this work, we combine the recent advances in the field of nanophotonics\cite{luk2017hybrid} and the non-paraxial theory of structured light\cite{das2015beam,lamprianidis2018excitation} to show that it is possible to circumvent the above-mentioned obstacles and design a non-paraxial cavity of high finesse with a subwavelength particle mirror. The cavity has high quality eigenmodes described by the even order magnetic multipole solutions of Maxwell’s equations, with the fundamental mode being produced by a longitudinal magnetic quadrupole. By utilizing anapole excitation, we design a subwavelength core-shell spherical particle with negligible dipole contributions at the magnetic quadrupole resonance. This core-shell spherical particle acts as a mirror with an almost prefect reflectivity. 

\begin{figure}
    \includegraphics[width=\linewidth]{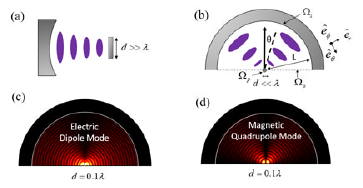}
    \caption{(a) Paraxial cavity with one mirror much larger than the wavelength. The fields are almost evenly distributed inside the cavity. (b) Non-paraxial cavity with a hemispherical shell mirror and a subwavelength mirror. Field amplitudes are highest on the small mirror. (c) Cavity mode for the case that the small scatterer scatters like an electric dipole(ED). (d) Cavity mode corresponding to the case that the small mirror scatters like a magnetic quadrupole(MQ).}
    \label{fig:my_label}
\end{figure}

Fig. 1(b) shows the geometry of the considered cavity. It consists of a large hemispherical shell and a small spherical particle that can be trapped. Unlike the paraxial case in Fig. 1(a), the size of the small mirror can be much smaller than the wavelength. Because of large difference between the radii of the two mirrors, it is expected that most of the energy inside the cavity is “focused” at the small particle mirror whereas, for a paraxial cavity, the energy is almost evenly distributed within the cavity. To confirm this, we first study the eigenmodes of the system.

The eigenmodes are solutions of the vectorial Helmholtz equation with proper cavity boundary conditions. In Fig. 1(b), the boundaries of the outer shell $\Omega_s$ and the small sphere $\Omega_p$ are assumed to be perfect electric conductor (PEC). We first assume a PEC boundary on $\Omega_o$ and later relax this requirement to obtain a low loss solution of the open cavity.


The eigenmode solutions are constructed using vector spherical harmonics(VSHs). We do not consider solutions with  azimuthal dependence because of symmetry. It should be emphasized that the set of all VSHs are linearly independent on the entire sphere but not on the hemisphere. On the hemisphere, the VSHs with even degree n are a muthually orthogonal set and likewise the VSHs with odd n. We therefore consider linear combinations of either only even or only odd modes to describe the electric fields within the cavity\cite{mishchenko2002scattering}:
\begin{equation}
   \vec E(kr,\theta)  = \sum\limits_n {{a_n}{{\vec M}_n} + {b_n}{{\vec N}_n}} {\text{ with }}n={\text{even or odd}}  
\end{equation}
 where $ {{\vec M}_n}  = {\gamma _n}{E_\phi }\hat \phi$ and ${{\vec N}_n}  = {\gamma _n}({E_r}\hat r + {E_\theta }\hat \theta ) $ are the electric fields produced by the magnetic and electric multipoles of degree $n>1$, and ${{\gamma }_{n}}=\sqrt{\frac{2n+1}{2\pi n(n+1)}}$ is the normalization factor. ${{E}_{r}}=\frac{n(n+1)}{kr}{{z}_{n}}(kr){{P}_{n}}(\cos \theta )$ is the radial component of the electric field and ${{E}_{\theta }}=\frac{1}{kr}\frac{d}{d(kr)}{{z}_{n}}(kr)\frac{d}{d\theta }{{P}_{n}}(\cos \theta )$, ${{E}_{\phi }}={{z}_{n}}(kr)\frac{d}{d\theta }{{P}_{n}}(\cos \theta )$ are the two angular components, and ${{z}_{n}}(kr)$ is a linear combination of the spherical Hankel function of the first and the second kind, $h_n^{(1)}(kr)$ and $h_n^{(2)}(kr)$, describing outgoing and incoming spherical waves respectively, with $k=2\pi/\lambda$ being the wavenumber in vacuum. ${{P}_{n}}(\cos \theta )$ is the Legendre polynomial of degree n. 
 
We then apply PEC boundary conditions to Eq. (1):
\begin{equation}
\left\{ \begin{matrix}
   {{E}_{\theta }}=0  \\
   {{E}_{\phi }}=0  \\
\end{matrix} \right.\text{ at }{{\Omega }_{s}}\text{ and }{{\Omega }_{p}}\text{, }\left\{ \begin{matrix}
   {{E}_{r}}=0  \\
   {{E}_{\phi }}=0  \\
\end{matrix} \right.\text{ at }{{\Omega }_{o}}
\end{equation}

The boundary conditions at ${{\Omega }_{s}}$ and $\Omega_p$ set up the standing wave along the radial direction, which can be described by the spherical bessel function $j_n(kr)$. We set the surface of the particle to be at $r=0$ to satisfy the boundary condition at $\Omega_p$. The length of the cavity along the radial direction is thus determined by the resonant condition, i.e. root of $j_n(kL)=0$ where $L$ is the cavity radius. This resonant condition requires the smallest cavity radius on the order of wavelength $\lambda$. For $n=1$, $L\approx0.715\lambda$ and $n=2$, $L\approx0.917\lambda$.  
%
The boundary condition at ${{\Omega }_{o}}$ demands special attention, it sets the condition for the standing wave pattern along the $\theta$ direction. Interestingly, this standing wave condition is solely determined by the properties of ${{P}_{n}}(\cos \theta )$  and has no dependence on the wavenumber $k$. Physically, the number of nodes of the standing wave along $\theta$ corresponds to the order of multipolar radiations such as dipole with $n=1$, quadrupole with $n=2$. It shows the possibility of having higher-order multipolar resonances inside a particle with diameter that is much smaller than the wavelength used as we will show in the following.

We look at the electric field of the electric and magnetic multipoles ${{\vec{N}}_{n}}$ and  ${{\vec{M}}_{n}}$ at ${{\Omega }_{o}}$ separately. For the magnetic multipoles ${{\vec{M}}_{n}}$, the electric field has only ${{E}_{\phi }}$ component. The boundary condition  ${{E}_{\phi }}=0$ at ${{\Omega }_{o}}$ is satisfied by the magnetic multipole solutions if n is even, 
with the lowerest order being the magnetic quadrupole(MQ)(n=2). In contrast, for the electric multipoles ${{\vec{N}}_{n}}$, $E_r$ is the only tangential component on $\Omega_0$ and to make it vanish n should be odd, with the lowest order being electric dipole(ED)(n=1).  

The modulus of the electric field of the ED and MQ eigenmodes are shown in Fig. 1(c) and Fig. 1(d). As predicted, both eigenmodes have a large field concentration at the small mirror. The $E_\theta$ component of the ED mode 
has maximum amplitude on ${{\Omega }_{o}}$. In fact, this property is shared by all electric multipoles of odd order. Therefore, if we remove the PEC boundary at ${{\Omega }_{o}}$ to open the cavity, this mode will strongly leak into the exterior causing large diffraction losses. For the even magnetic multipoles, the situation is very different. As can be seen in Fig. 1(d), the electric field of the MQ mode is zero at ${{\Omega }_{o}}$ and the same behavior is observed for all other even order magnetic multipoles. In this case, removing the PEC boundary on ${{\Omega }_{o}}$ has much less influence. If we remove the PEC boundary at ${{\Omega }_{o}}$ to obtain a half open cavity with only the small particle present, the field
\begin{equation}
\begin{split}
{{\vec E}_n}(kr,\theta ) & = {{\vec M}_n}(kr,\theta ) \\ 
   & = {\gamma _n}{j_n}(kr)\frac{d}{{d\theta }}{P_n}(\cos \theta )\hat \phi {\text{ with }}n = 2,4... 
\end{split}
\end{equation}%
satisfies Maxwell's equations inside the open cavity and the PEC conditions on the outer  shell and the small particle mirror, while its field is small on $\Omega_0$. It therefore is an approximate eigenmode with only small diffraction losses\cite{supple}. As mentioned, these eigenmodes are the fields produced by the even order magnetic multipoles, with the lowest order being the field of the MQ. Instead of the electric dipole approximation when designing a mirror of subwavelength size,  this eigenmode description requires  to consider higher-order magnetic multipole excitations.

Now the question is how to construct the two mirrors of the cavity. The large curved mirror on the outer shell can be constructed using multi-layered Bragg structures by considering a plane wave approximation at each angle $\theta$.  The challenge is the small particle mirror. Since the particle is small, one cannot take the plane wave approximation to design such a mirror as for the larger mirror case. This problem is equivalent to designing a small particle that reflects nearly all the tightly focused eigenmodes given by Eq. (3), with $j_n(kr)$ replaced by $h_n^{(2)}(kr)$ from the upper hemisphere with $\theta \in [0,\frac{\pi}{2}]$ as shown in the inset of Fig. 2(a). 

We use generalized Mie theory to describe this interaction\cite{mishchenko2002scattering} and consider expanding the incoming and outgoing fields into magnetic multipoles $\vec{M}_n(kr,\theta)$. Since the field has only $E_{\phi}$ component, the total electric farfield can be written as:
\begin{equation}
\begin{split}
     E_{tot,\phi}^{far} &= E_{in,\phi}^{far} + E_{out,\phi}^{far} \\ 
   &= \sum\limits_{n = 1}^\infty  {A_n^{in}} \frac{{{e^{ - ikr}}}}{{kr}} + \sum\limits_{n = 1}^\infty  {A_n^{out}} \frac{{{e^{ikr}}}}{{kr}}  
\end{split}
\end{equation}
in which ${A_n^{in}}=\frac{1}{2}{a_n}{(i)^{n + 1}}{M_n}(\theta)$, ${A_n^{out}}=(\frac{1}{2}{a_n}+p_n){(-i)^{n + 1}}{M_n}(\theta)$ are coefficients that determine the angular distributions of the incoming and outgoing spherical waves over $\theta \in [0,\pi]$ respectively. $M_n(\theta)=\frac{d}{d\theta}P_n(\text{cos}\theta)$ corresponds to the angular distributions of the magnetic multipoles. The incident field coefficients $a_n$ and the scattered field coefficients $p_n$ are related by the transition matrix $p_n=T_na_n$.

The incoming spherical waves $E_{in,\phi}^{far}$ has an angular distribution that is non-zero for  $\theta \in [0,\frac{\pi }{2}]$ and zero for  $\theta \in [\frac{\pi }{2},\pi ]$. In order to have the particle to reflect all of the incoming field, the outgoing spherical wave $E_{out,\phi}^{far}$  should have the same angular distribution as the incoming one over $\theta \in [0,\pi ]$. This is fulfilled by letting $A_n^{in}=\pm A_n^{out}$. Since ${{p}_{n}}={{T}_{n}}{{a}_{n}}$, we obtain the following requirements: 
\begin{equation}
    {T_n} = \{ \begin{array}{*{20}{c}}
  { - 1\text{\;for\;n\;even\;}} \\ 
  {0\text{\;for\;n\;odd}} 
\end{array}\;or{\text{   }}{T_n} = \{ \begin{array}{*{20}{c}}
  {0\text{\;for\;n\;even\;}} \\ 
  { - 1\text{\;for\;n\;odd}} 
\end{array}
\end{equation}
which means that half of the magnetic multipoles should be at resonance with ${{T}_{n}}=-1$ and the other half of different parity should be zero. This can be understood from symmetry considerations. The particle scatters incident light into different multipoles. To have the particle reflect all the incident light, the scattered light pattern should be symmetrical vertically with respect to the focus to interfere destructively in the lower hemisphere($\theta \in [\frac{\pi }{2},\pi ]$) and constructively in the upper hemisphere($\theta \in [0,\frac{\pi }{2} ]$) with the incident light. If one looks at the parity of different multipoles, they are different depending on whether n is even or odd. It is well known that for a particle to have strong directional scattering, the particle should scatter multipoles of different parities evenly. However, in the case of total reflection, only multipoles of the same parity should remain to keep the symmetry of the scattering pattern. We consider the even half of the magnetic multipole resonances as they are the low-loss modes for the cavity. It is hard to design a particle that resonates at all the even order magnetic multipoles, while  the contributions from all the odd magnetic multipoles are zero. Things can be greatly simplified if the illumination and the particle are chosen with only the first few dominant multipole terms. 

Let$'$s consider the case that the illumination has the same angular distribution as the MQ term $M_2(\theta)$. If one performs a mode decomposition of this mode into different multipoles in the upper hemisphere, one can see that the other even magnetic multipoles have zero amplitude, while there is an infinite number of odd modes with non-zero coefficients as required by the orthogonality property of the Legendre polynomials in the hemisphere.

If the particle has in addition to the MQ contribution, also other odd magnetic multipole contributions, they can be excited as well. Because they have different parities than the MQ, the scattering pattern is no longer symmetric, and this prevents all the light being reflected.

What we need is a particle that has a pure magnetic quadrupole resonance, and an incident incoming field shaped into the $M_2(\theta)$ angular distribution. For a small particle, the leading term is the magnetic dipole term. Its contribution can be eliminated by the anapole condition using core-shell particle design as we show in the following.

We first look at a case of a silicon sphere of radius $r$ with a refractive index n=3.5 under  ${{{M}}_{2}(\theta)}$ illumination. The reflectivity  is shown in Fig. 2(a). 
\begin{figure}
    \includegraphics[width=\linewidth]{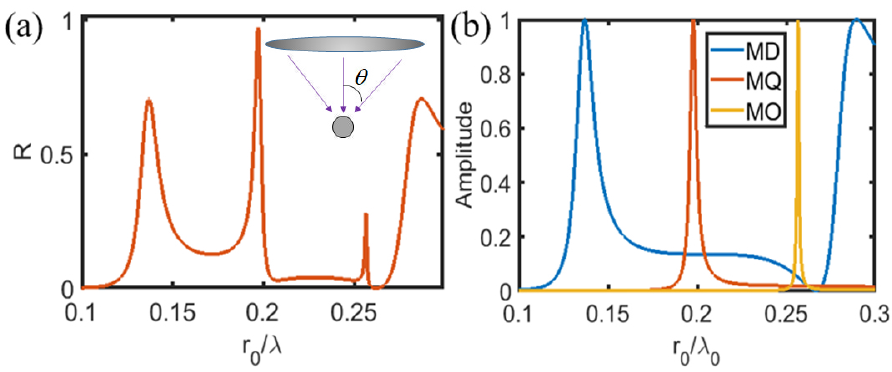}
    \caption{(a) Reflectivity from a silicon sphere with n=3.5 under magnetic quadrupole field illumination. (b) Multipolar contribution to the T-matrix from the silicon sphere (MD=magnetic dipole, MQ=magnetic quadropole and MO=magnetic octopole).}
    \label{fig:my_label}
\end{figure}
There are several reflection peaks with the highest $R\approx 0.97$ for radius ${r}\approx 0.18\lambda $. A multipolar decomposition of the particle’s T-matrix in Fig. 2(b) reveals that this peak is due to the MQ resonance. However, it can be seen that at the MQ resonance, there is also a non-zero MD contribution, which is undesired.

At ${r}\approx 0.27\lambda $, the contribution of the MD is almost zero. This special point is related to the anapole condition of the MD\cite{luk2017hybrid,lamprianidis2018excitation}. Because the Cartesian dipole and toroidal dipole have the same radiation patterns but opposite phase, the net scattering power from the dipole term is zero in the far-field. Normally this point is used for the cloaking of small particles when the dipole term is dominant. If this point can be tuned such that it coincides with the MQ resonance, we can excite a pure MQ and this, instead of cloaking, will produce a better mirror.

To show this, we consider a $\text{Si@SiO}_2$ core-shell spherical particle. We vary the particle's inner and the outer radii $r_1$ and $r_2$, and calculate the contributions from the MD, ${{T}_{1}}$ (Fig. 3(b)) and MQ, ${{T}_{2}}$ (Fig. 3(a)). 
At ${{r}_{1}}=0.18\lambda $ and ${{r}_{2}}=0.18\lambda $, the anapole overlaps with the MQ resonance. We mark this point as the perfect mirror point. At this point, the ratio between  ${{T}_{1}}$ and ${{T}_{2}}$ is almost zero as shown in Fig. 3(c). In Fig. 3(d), we plot the reflectivity of this particle under $M_2$ illumination. A very high $R=99.993\%$ is at the perfect mirror point. Assuming a similar reflectivity for the large mirror, the cavity with this $\text{Si@SiO}_2$ particle mirror is expected to have a much higher finesse $\mathcal{F} = \pi R^{1/2}/(1-R)\approx4.5\times10^4$ than $\mathcal{F} \approx 103$ for the  silicon sphere with $R\approx 97\%$.

\begin{figure}
    \includegraphics[width=0.95\linewidth]{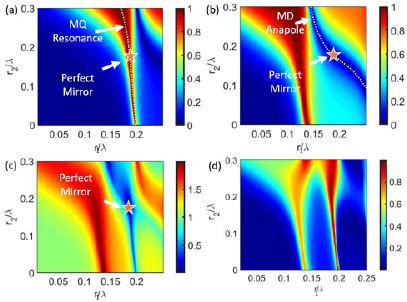}
    \caption{Contributions of the different magnetic multipole components to the T-matrix of the of a Si-core SiO2-shell sphere as a function of inner and out radius. (a) $T_2$ from MQ component. (b) $T_1$ from MD component. (c) The ratio  of $T_1$ and $T_2$. (d) The total reflectivity of the sphere under $M_2$ illumination.}
    \label{fig:my_label}
\end{figure}

We use Comsol to rigorously compute the eigenmode of the open cavity as a validation of the cavity mode solutions obtained. The inner particle is replaced by the core-shell particle while $\Omega_s$ is still PEC boundary.  The boundary $\Omega_o$ is removed and the whole cavity is surrounded by the perfect matched layers. The mode found by Comsol shown in Fig. 4(a) is indeed very similar to the MQ mode stated above.

In addition, we show the Q-factor of the cavity for different cavity length L in Fig. 4(b). For comparison, we also include the case where the small particle ($n=5, r=0.135\lambda$) has a pure ED resonance, which corresponds to a cavity with the ED mode. The Q-factor increases as the cavity length increases, and the Q-factor difference between the MQ mode and the ED mode is huge, almost an order of magnitude\cite{supple}.

We note that the particle mirror can be very small as long as it has a pure even order magnetic resonance. In Fig. 4(c), we make a perfectly reflecting "artificial  atom" with diameter less than $0.1\lambda$. The particle provides efficient subwavelength focusing, because there is a huge field enhancement around the particle\cite{supple}. With the particle being replaced by an atom, the weak multipole resonances can be greatly enhanced\cite{mizushima1964delta,alaee2020quantum}.

\begin{figure}[hpt]
 \includegraphics[width=0.95\linewidth]{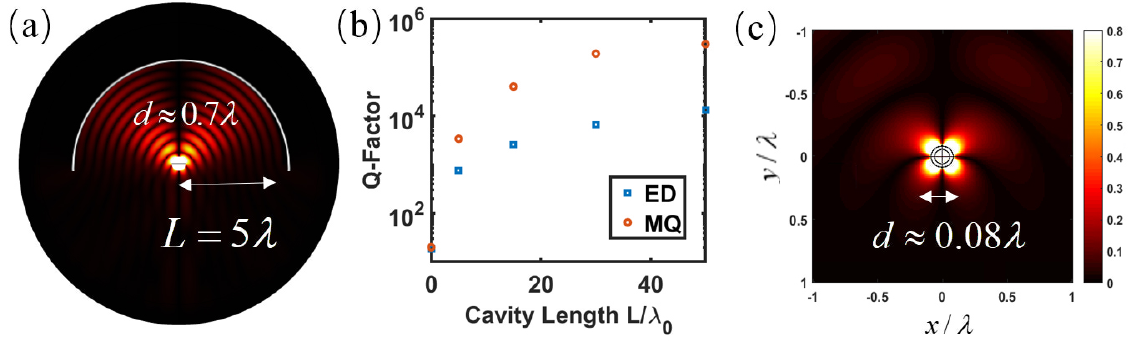}
    \caption{(a) Eigenmode profile obtained with Comsol using the core-shell particle as the small mirror. (b) Q-factor for ED mode and MQ mode for different cavity length L. (c) Nearfield distribution of a totally-reflecting core-shell spherical particle with a diameter less than $0.1\lambda$. The refractive index of the inner core is 20 and the outer-shell is 8.6.}
    \label{fig:my_label}
\end{figure}



As a final remark, we consider application of our cavity system in the field of levitated optomechanics\cite{millen2020optomechanics,barker2010cavity,chang2010cavity,romero2010toward},. We estimate the single-photon coupling rate $g_0$, which is of fundamental importance as it quantifies the interaction strength between a single photon and a single phonon\cite{millen2020optomechanics}. The estimated $g_0$ in our system is on the order of tens of GHz\cite{supple}.  This is a significant improvement over the typical value of tens of Hz in the field of levitated optomechanics where a non-resonant Rayleigh particle is used\cite{romero2011optically}. Considering the high mechanical quality factor due to levitation and the high finesse of the cavity, the system is expected to have interesting applications in the strong optomechanical coupling region.

In conclusion, we have shown that one can design a high finesse cavity with a subwavelength particle mirror. The key idea is to excite a pure magnetic quadrupole resonance with the help of anapole excitation. Due to the subwavelegnth size of the particle mirror and the high finesse of the cavity, this system is expected to have important applications in optomechanical sensing, nano-photonics, lasing, nonlinear optics and the exploration of the higher order multipolar light-matter interactions.

\section*{Acknowledgement}
The authors would like to thank Y.M. Blanter and S. Gr\"oblacher for their helpful discussions on the potential applications in the field of quantum optomechanics.

\end{document}